\documentstyle [aps,prc,preprint]{revtex}
\begin {document}
\draft
\preprint{SUNY-NTG-96-42} 
\title
{The Two-Pion Exchange {\em NN} Potential in Nuclear Matter and Nuclear
 Stability}

\author
{R. Rapp$^{1,2}$, J.W. Durso$^{1,2,3}$, and J. Wambach$^{4}$}
 
\address
{1) Inst. f\"{u}r Kernphysik, Forschungszentrum
     J\"{u}lich GmbH, \\ D-52425 J\"{u}lich, Germany \\
 2) Dept. of Physics, SUNY at Stony Brook, Stony Brook NY 11794-3800,
    U.S.A. \\
 3) Physics Department, Mount Holyoke College, South Hadley,
    MA 01075, U.S.A.\\
 4) Inst. f\"ur Kernphysik, Schlo{\ss}gartenstr.~9, TH-Darmstadt, 
    D-64289 Darmstadt, Germany}
 
\maketitle
 
\begin{abstract} 
A meson exchange model of the $\pi\pi$ interaction which fits 
free $\pi\pi$ scattering data is used to calculate the interactions
of pions in nuclear matter as a function of nuclear density.
Polarization of the nuclear medium by the pions results in a 
marked increase in the s-wave $\pi\pi$ attraction at low
energy.  The influence of this effect on the nucleon-nucleon
interaction is a corresponding increase with density of the
$NN$ central potential due to the exchange of two correlated
pions, resulting in an $NN$ interaction which fails to 
saturate.  A possible mechanism for restoring the theoretical
stability of nuclear matter is explored and found to be effective.

\end{abstract}
\pacs{ } 

\newpage

\section{Introduction} 
At the level of effective interactions between hadrons, the pion and 
the $\pi\pi$ interaction occupy a central position.  For example, 
in the two-nucleon system, single-pion 
exchange provides the longest-range part of the interaction, 
and two-pion exchange is the agent of the intermediate-range 
central attraction~\cite{MHE,Mach} which is necessary for nuclear 
binding.  However, in order to provide sufficient attraction, 
it is essential that the exchanged pions be {\em correlated}; 
that is, that their ``in-flight'' interaction be taken into account.  
This, in turn, requires a reliable model for the
$\pi\pi$ interaction, the development of which has been the subject 
of considerable effort~\cite{LDHS,PHS,JPHS,ARCSW,RDW}.

Hand in hand with the the development of a reliable model of the 
$\pi\pi$ interaction has come the effort to understand what effect 
the presence of a nuclear medium has on the interaction.  This 
effect is of interest  for its consequent influence 
on the nucleon-nucleon interaction in the presence of other 
nucleons, as well as, for example, in the analysis of 
heavy-ion collisions, where it may play 
a significant role in explaining the spectrum of dilepton 
production~\cite{CRW}.  Our focus here will
be on the modification of the nucleon-nucleon interaction.

Let us begin by summarizing briefly and qualitatively the results 
of our and others' work on the $\pi\pi$ interaction and its 
modification by the nuclear medium.  The J\"ulich Model~\cite{LDHS,PHS}
for the free $\pi\pi$ interaction gives a quantitatively accurate
description of the interaction in low partial waves over a broad 
energy range.  
Subsequent investigations~\cite{CASN,MWS,ACS} of the effect of the 
medium on the $\pi\pi$ interaction through the coupling of pions 
to nucleon-hole and $\Delta$-hole configurations revealed that 
almost all of the models typically used
for the $\pi\pi$ system led to a considerable build-up of attractive 
strength near and below the physical $\pi\pi$ energy threshold in 
the s-wave ($\sigma$ channel), resulting in $\pi\pi$ bound states 
at subnuclear density. A preliminary estimate~\cite{DKW} of the 
effect of the redistribution of strength on the in-medium 
nucleon-nucleon interaction indicated only minor alterations of the
strength and range of the attraction due to correlated 2$\pi$ exchange. 
Missing from this estimate, however, were 
the effect of the sub-threshold strength and an accurate treatment 
of the 2$\pi$ propagator.

Further investigation of the medium modification of the $\pi\pi$
amplitude revealed a great sensitivity~\cite{RDW} to the particular 
scattering equation used to calculate it---that is, 
to the way in which the interaction is evaluated off-shell.  
Indeed, the original J\"ulich Model~\cite{LDHS,PHS} extrapolated to 
the nuclear medium displayed a $\pi\pi$ pairing instability ({\em 
i.e.}, the formation of $\pi\pi$ bound states of less than zero 
c.m. energy) at only slightly above normal nuclear matter 
density~\cite{ARCSW}.  
This difficulty was eventually traced to a failure of the interaction 
in the J\"ulich Model to satisfy low-energy theorems of chiral 
symmetry, {\em e.g.} that the s-wave scattering lengths did not vanish 
in the limit of zero pion mass.  An ``improved'' model~\cite{RDW}, 
which corrected this deficiency, yet preserved
the good description of the $\pi\pi$ data, coupled with a kinematic 
prescription for continuing the $\pi\pi$ potential 
off-shell which preserves the scattering length, succeeded in 
removing the instability or, at least, in delaying its onset
to much higher density.

In this report we will investigate how our best model to date of the 
$\pi\pi$ interaction, extended to take the nuclear medium into 
account, affects the nucleon-nucleon force.  In doing this we 
will be concerned especially with the implications of the 
density-dependence that we find in the force  
for the stability of nuclear matter.  We shall see that 
the conventional methods that we have applied lead to serious 
problems for the saturation of nuclear matter, and we shall 
explore one unconventional method which could be 
a remedy for the problem.

\section{Dynamical Models and Numerical Results}
\subsection{Meson Exchange Model for $N\bar N\to\pi\pi$
and Corresponding $NN$ Potential in Free Space} 
For the evaluation of the correlated two-pion exchange contribution 
to the low-energy nucleon-nucleon interaction we adopt the recent 
approach of the Bonn-J\"ulich group. It is based on a meson exchange 
model for the $N\bar N\to\pi\pi$ reaction which by means of a 
dispersion relation is then transformed to the $NN$ channel. The 
$N\bar N\to\pi\pi$ model in free space has been constructed in 
refs.~\cite{KDH,SHSPD} for the so-called pseudophysical region, 
{\it i.e.,} well below the physical nucleon-antinucleon threshold, 
which is the relevant regime for extracting a $NN$ potential.  
The corresponding scattering amplitude, $\tau_{N\bar N\to\pi\pi}$, 
is composed of a transition Born amplitude $\tau_{B,N\bar N\to\pi\pi}$ 
(consisting of nucleon and $\Delta$ exchange), and a rescattering term 
accounting for the $\pi\pi$ final state interaction. Within the 
Blankenbecler-Sugar (BbS) reduction scheme~\cite{BbS} of the 
underlying relativistic equation, and after a partial wave 
expansion, the scattering equation takes the form     
\begin{eqnarray}   
\lefteqn{ 
\langle 00|\tau_{N\bar N\to\pi\pi}^{JI}(t',q,p)|
\lambda_N\lambda_{\bar N} \rangle  =    
 \langle 00|\tau_{B,N\bar N\to\pi\pi}^{JI}(t',q,p)|
\lambda_N\lambda_{\bar N} \rangle    } 
   \nonumber\\ 
 & & + \int\limits_0^\infty\frac{k^2 dk}{(2\pi)^2} \  
\langle 00|M_{\pi\pi}^{JI}(t',q,k)|00 \rangle \ 
G_{\pi\pi}^0(t',k) \ \langle 00|\tau_{B,N\bar N\to\pi\pi}^{JI}
(t',k,p)| \lambda_N\lambda_{\bar N} \rangle \ .   
\label{tau}
\end{eqnarray}  
where $\lambda_N$ and $\lambda_{\bar N}$ denote the helicities of 
nucleon and antinucleon, $p$ the magnitude of their c.m.~momenta, and 
$q$ the c.m.~momentum of the two pions. 
$t'\equiv E^2$ is the total energy
squared of the $N\bar N$ ($\pi\pi$) system, but it will play the 
role of a t-channel 4-momentum transfer once the analytic continuation 
to the $NN$ system is performed. Conservation of $G$-parity forces 
the spin-isospin quantum numbers $JI$ to satisfy $J$+$I$=even. The  
$\pi\pi$ invariant scattering amplitude will be discussed in more 
detail below. The uncorrelated two-pion propagator in free space reads  
\begin{equation} 
G_{\pi\pi}^0(t',k)=\frac{1}{\omega_k} \ \frac{1}{t'-4\omega_k^2
+i\epsilon} 
\end{equation} 
with $\omega_k^2=m_{\pi}^2+k^2$. The $N\bar N \to \pi\pi$ transition 
amplitude (and the $\pi\pi$ interaction) also contains the coupling to 
intermediate $K\bar K$ states, which are not explicitly written in 
eq.~(\ref{tau}), but they are included in all our calculations. The 
relation of the $\tau$ amplitude to the in the literature frequently 
used Frazer-Fulco helicity amplitudes $f_\pm^{J}$~\cite{FF} is given by 
\begin{eqnarray}  
f_+^{JI}(t') & = & \frac{p_{on}M_N}{4 (2\pi)^2 (p_{on}q_{on})^J} 
 \ \langle 00|\tau_{N\bar N\to\pi\pi}^{JI}(t',q_{on},p_{on})| 
\frac{1}{2}\frac{1}{2}\rangle 
\nonumber\\   
f_-^{JI}(t') & = & -\frac{p_{on}M_N}{2 (2\pi)^2 \sqrt{t} 
(p_{on}q_{on})^J} \ \langle 00|\tau_{N\bar N\to\pi\pi}^{JI}
(t',q_{on},p_{on})| \frac{1}{2}(-\frac{1}{2})\rangle 
\label{ffamp}
\end{eqnarray}  
where the on-shell momenta are defined by 
\begin{eqnarray}  
p_{on} & = & \left(\frac{t'}{4}-M_N^2\right)^{1/2} 
\nonumber\\
q_{on} & = & \left(\frac{t'}{4}-m_\pi^2\right)^{1/2} \ .  
\end{eqnarray}    

In refs.~\cite{KDH,SHSPD} the J\"ulich $\pi\pi$ meson exchange 
model~\cite{LDHS,PHS} has been employed for the $\pi\pi$ amplitude 
entering eq.~(\ref{tau}). This model gives a satisfactory description 
of the available $\pi\pi$ scattering data up to rather high energies
(beyond 1~GeV) and partial waves up to $J$=2. The 
$N\bar N \to \pi\pi$ model is 
then completed by adjusting the four free parameters of the transition 
Born amplitudes $\tau_{B,N\bar N\to\pi\pi}$ 
(two form factor cutoffs and 
couplings for N and $\Delta$ exchange; see ref.~\cite{SHSPD} for 
details) to reproduce the quasi-empirical Frazer-Fulco amplitudes
in the pseudophysical region $4m_{\pi}^{2} \le t < 4M_N^2$ ~\cite{Hoeh}.
 (See the left column of fig.~1.)\\ 

However, as mentioned earlier, the application of the J\"ulich  
model to the calculation   
of s-wave $\pi\pi$ correlations in nuclear matter has been shown 
to result in an unrealistic $\pi\pi$ pairing instability (pion pair 
condensation) slightly above saturation density~\cite{ARCSW}. As we 
will see in the next section, such a behavior indeed drastically 
increases the attraction in the intermediate and long range of the 
$NN$ potential in the nuclear medium, certainly contradicting any  
mechanism of nuclear saturation. On the other hand, as has been 
demonstrated in refs.~\cite{ARCSW,RDW}, the implementation of 
constraints dictated by chiral symmetry in the $\pi\pi$ interaction 
significantly suppresses the tendency towards pair condensation. To 
investigate more quantitatively how the in-medium properties of the 
central $NN$ potential are affected by the chiral constraints in 
the $\pi\pi$ sector, we will also perform calculations with a chirally 
improved version of the J\"ulich $\pi\pi$ interaction~\cite{RDW}. 
These ``minimal'' chiral improvements consist of 
\begin{itemize} 
\item[(i)] the introduction of $\pi\pi$ contact interactions as implicit
in the gauged non-linear $\sigma$ model of Weinberg~\cite{We66}; they 
ensure that the tree level $\pi\pi$ amplitude satisfies the soft pion 
theorems; in particular, the contact terms constitute a strong source
of subthreshold repulsion in the $\pi\pi$ interaction kernel; 
\item[(ii)] a slight modification in the off-shell continuation of  
the interaction kernel of the scattering equation: rather than using 
the (on-energy-shell) BbS prescription we employ an on-mass-shell 
prescription; this ensures that the fully iterated $\pi\pi$ amplitude     
preserves the correct chiral limit for the s-wave scattering lengths 
({\it i.e.}, $a^{0I}\to 0$ for $m_\pi \to 0$). 
\end{itemize}   
A slight refit of the model parameters gives a similarly good 
description of the free $\pi\pi$ scattering data to that of the 
original J\"ulich model, while the $\pi\pi$ pair condensation in 
nuclear matter is delayed to much higher density---about 
2$\rho_0$ ~\cite{RPhD}. As before, the additional four free 
parameters in the $N\bar N$ sector are adjusted to reproduce the 
quasi-empirical Frazer-Fulco amplitudes, {\em cp.} the right column in
fig.~1 (for consistency, the off-shell 
continuation of the corresponding transition Born amplitudes 
$\tau_{B,N\bar N\to\pi\pi}$ is also changed to on-mass-shell). 

In the rest of  this section we will discuss the procedure 
for extracting a central $NN$ potential due to correlated two-pion 
exchange in the $\sigma$ channel. Following refs.~\cite{DKW,KDH}, the 
relevant spectral function is related to the Frazer-Fulco 
amplitudes by 
\begin{equation}
\eta_{00}(t')=24\pi \sqrt{\frac{t'-4m_\pi^2}{t'}} 
\frac{1}{(t'-4M_N^2)^2} \left[ |f_+^{00}(t')|^2-
|f_{B,+}^{00}(t')|^2 \right] \Theta(t'-4m_\pi^2) \ , 
\label{eta00f} 
\end{equation} 
where the subtraction of the Born amplitude $|f_{B,+}^{00}|^2$ 
removes the iterated single-pion exchange part of the $2\pi$ exchange. 
The $NN$ potential in momentum space is then obtained by means 
of a dispersion integral along the unitarity cut, the latter 
starting at the free two-pion threshold: 
\begin{equation}  
V_{NN\sigma}(t)=-\frac{\kappa}{\pi} \int\limits_{4m_\pi^2}^{\infty}
dt' \ \frac{\eta_{00}(t')}{t-t'} \ P_1, 
\label{VNNt} 
\end{equation} 
with $\kappa=M_N^2/[(2\pi)^3(E_{p_1}E_{n_1} E_{p_2}E_{n_2})^{1/2}]$
and the nucleon energies $E_{p_1}$=$E_{n_1}$= $\sqrt{M_N^2+k^2}$,
 $E_{p_2}$=$E_{n_2}$= $\sqrt{M_N^2+k'^2}$ ({\em cp.} fig.~2).  
The operator $P_1=1^n 1^p$ is to be taken between the spinors of 
the incoming and outgoing nucleons. As suggested in ref.~\cite{DKW}, 
a more intuitive representation of our results in the coming 
section is obtained by a Fourier transformation of the momentum 
space potential into coordinate space. In the quasistatic limit 
one has $t\simeq-(\vec k'-\vec k)^2\equiv\vec k''^2$ and 
$\kappa=1$, so that 
\pagebreak
\begin{eqnarray}  
V_{NN\sigma}^C(r) & = & -\frac{1}{\pi} \int \frac{d^3k''}{(2\pi)^3} 
 \ e^{i\vec k''\cdot \vec r} \int\limits_{4m_\pi^2}^{t_c} dt' \ 
\frac{\eta_{00}(t')}{t-t'} \nonumber\\    
     & = & -\frac{1}{\pi}\int\limits_{4m_\pi^2}^{t_c} dt' \ 
\frac{\eta_{00}(t')}{4\pi} \ \frac{\exp (-\sqrt{t'}r)}{r} \ , 
\label{vnnsr}
\end{eqnarray} 
which is simply a superposition of Yukawa-type potentials with a 
continuous mass/coupling constant distribution $\eta_{00}(t')$. 
Cutting off the integration at    
at $t_c=50m_\pi^2$ avoids the extrapolation of our model to physically 
unreasonable momentum transfers. Due to the presence of the 
exponential, our numerical results are practically insensitive to 
moderate variations in $t_c$. 

However, in carrying eqs.~(\ref{eta00f}) and (\ref{vnnsr}) over 
to the in-medium case the following problem arises: the unitarity cut 
of the 2$\pi$ continuum no longer starts at the free 2$\pi$ 
threshold ($E=\sqrt{t'}=2m_\pi$) but moves all the way down to 
$E$=0; this is due to the dressing of the pions with $NN^{-1}$ and 
$\Delta N^{-1}$ excitations in nuclear matter, which generates an 
imaginary part of the two-pion propagator~\cite{ACS}, 
\begin{equation}    
Im G_{\pi\pi}(E,k)=-\int\limits_{0}^{E} \frac{d\omega}{\pi} \ 
Im D_\pi(\omega,k) \ Im D_\pi(E-\omega,k) \ , 
\end{equation}  
extending below $E$=2$m_\pi$. Consequently, $Im M_{\pi\pi}$ 
also becomes nonzero below threshold. On the other hand, an 
extension of eq.~(\ref{eta00f}) below $t'=4m_\pi^2$ is not obvious 
since already the {\it free} Frazer-Fulco amplitudes do not vanish 
below threshold. In ref.~\cite{DKW} the subthreshold contribution
to the medium modified $\eta_{00}$ was neglected  
by employing eq.~(\ref{eta00f}) with in-medium $f^0_+$ amplitudes. 
However, since we intend to study the impact of in-medium 
$\pi\pi$ effects such as possible bound state 
formation on the $NN$ potential, the 
subthreshold strength in the $\pi\pi$ amplitude has to be 
accounted for. This can be achieved by formulating a scattering 
equation for $N\bar N \to \pi\pi \to N\bar N$ and relating the 
imaginary part of $M_{N\bar N}$ in the pseudophysical region 
to $\eta_{00}$. For an in- and 
outgoing $N\bar N$ pair with momenta $\vec n,-\vec n$ and 
$\vec p,-\vec p$, respectively ({\it cp.} fig.~2), and total 
energy $\sqrt{t'}$ one obtains in the BbS framework: 
\pagebreak
\begin{eqnarray} 
\lefteqn{ M_{N\bar N}(t',\vec p,\vec n)  =  \frac{1}{2} \int
\frac{d^3q}{(2\pi)^3}  \ \tau_{B}^\dagger(t',\vec p,\vec q) 
 \ G_{\pi\pi}^0(t',q) \ \tau_{B}(t',\vec q,\vec n)}  \qquad \
\nonumber\\
& & + \frac{1}{2}\int\frac{d^3q}{(2\pi)^3} \ 
\tau_{B}^\dagger(t',\vec p,\vec q) \ G_{\pi\pi}^0(t',q) 
\int\frac{d^3q'}{(2\pi)^3} \ M_{\pi\pi}(t',\vec q, \vec q~') \ 
G_{\pi\pi}^0(t',q')  \ \tau_{B}(t',\vec q~',\vec n) \ , \quad  
\end{eqnarray}  
where $\tau_{B}^\dagger$ is the hermitian conjugate of 
$\tau_{B}$. The factors of $1/2$ account for the appearance of 
closed loops of identical bosons. A partial wave expansion leads to 
\begin{eqnarray} 
\langle \Lambda|M_{N\bar N}^{JI}(t')|\Sigma\rangle & = & 
\frac{1}{2} \int\limits_0^\infty\frac{q^2dq}{(2\pi)^3} \ 
\langle 00|\tau_{B}^{JI}(t',q,p)|\Lambda\rangle^* \ G_{\pi\pi}^0(t',q) 
 \ \langle 00|\tau_{B}^{JI}(t',q,n)|\Sigma\rangle  
\nonumber\\
 &  & + \frac{1}{2}\int\limits_0^\infty\frac{q^2dq}{(2\pi)^3} \ 
\langle 00|\tau_{B}^{JI}(t',q,p)|\Lambda\rangle^* \ G_{\pi\pi}^0(t',q) 
\nonumber\\ 
 & & ~~~~~~~~\times
\int\limits_0^\infty\frac{q'^2dq'}{(2\pi)^2} \ 
M_{\pi\pi}^{JI}(t',q,q') \ G_{\pi\pi}^0(t',q') \ 
\langle 00|\tau_{B}^{JI}(t',q',n)|\Sigma\rangle  
\nonumber\\
 & \equiv & \langle \Lambda|M_{N\bar N}^{JI,bare}(t')| \Sigma\rangle 
 \ + \ \langle \Lambda|M_{N\bar N}^{JI,rescat}(t')| \Sigma\rangle
\label{mnnbji}
\end{eqnarray}  
with $\Sigma=\sigma\bar\sigma$, $\Lambda=\lambda\bar\lambda$ 
denoting the $N\bar N$ helicity states. For our purpose we need
the on-shell values of $M_{N\bar N}$ in the pseudophysical region,  
therefore $n=p=\sqrt{t'/4-M_N^2}$ and consequently the transition 
Born amplitudes $\tau_B^{JI}$ are purely imaginary.  
Thus the combination $\tau_{B}^{JI}(t')^*\tau_{B}^{JI}(t')$ is 
always real, so that the imaginary part of $M_{N\bar N}$ is solely 
generated by on-shell intermediate $\pi\pi$ states.
Therefore, in free space, it vanishes below $\sqrt{t'}\le 2m_\pi$. 
On the other hand, in the nuclear enviroment an in-medium amplitude 
$Im M_{\pi\pi}$ that does not vanish below threshold naturally 
generates a corresponding nonzero 
$Im M_{N\bar N}(\sqrt{t'}\le 2m_\pi)$; {\it i.e.}, 
within the framework of the scattering eq.~(\ref{mnnbji}) the 
inclusion of the subthreshold region is well defined.   
It remains to relate $Im M_{N\bar N}$ to the spectral function 
$\eta_{00}$. From eqs.~(\ref{tau}),(\ref{ffamp}),(\ref{eta00f}),
(\ref{mnnbji}) one finds 
\begin{equation}  
Im \left[\langle ++|M_{N\bar N}^{00,rescat}(t')|++\rangle\right] = 
\frac{4\pi}{3} \ \eta_{00}(t') \ \langle ++|P_1|++\rangle \ .  
\end{equation}  
With 
\begin{eqnarray}  
\langle ++|P_1|++\rangle & = & (\bar u_\lambda(\vec p) 1^p 
v_{\bar\lambda}(-\vec p)) \  (\bar v_{\bar\sigma}(-\vec n) 1^n 
 u_\sigma(\vec n)) 
\nonumber\\
 & = & \frac{t'-4M_N^2}{4M_N^2} \ ,  
\label{spinME} 
\end{eqnarray} 
we arrive at  
\begin{equation} 
\eta_{00}(t') = \frac{4M_N^2}{t'-4M_N^2}\frac{3}{4\pi}   
Im \left[\langle ++|M_{N\bar N}^{00,rescat}(t')|++\rangle\right] \ ,  
\label{eta00m} 
\end{equation}  
which now allows the desired extension of $\eta_{00}$ below 
$\sqrt{t'}=2m_\pi$. 

\subsection{The $\sigma$ channel of the  $NN$ Potential  
in Nuclear Matter}
As suggested in ref.~\cite{DKW}, we restrict the medium effects in the 
$NN$ potential to be solely generated by the medium modifications in 
the $\pi\pi$ interaction $M_{\pi\pi}^{00}$ entering the rescattering 
term in 
eq.~(\ref{mnnbji}). Both the transition Born amplitude $\tau_B^{00}$  
and the explicitly-appearing two-pion propagator $G_{\pi\pi}^0$ 
remain in their vacuum form; the inclusion of medium effects in 
these two quantities would require further assumptions concerning the 
treatment of nucleons in the nuclear medium within a $NN$ scattering 
equation ({\it e.g.}, Bethe-Goldstone equation). Furthermore, we want 
to stay as close as possible to the spirit of the one-boson exchange 
picture, and compare the results to our earlier findings. 
The medium modifications of $M_{\pi\pi}^{00}$ are induced by dressing 
the intermediate pion propagators with the standard p-wave 
nucleon-nucleonhole ($NN^{-1}$) and delta-nucleonhole ($\Delta N^{-1}$) 
excitations~\cite{ErWe}. 
In refinement of the three-branch model utilized in 
refs.~\cite{MWS,DKW}, we here retain the full off-shell propagation 
dynamics of the pions in nuclear matter by means of a numerical 
treatment as described in detail in refs.~\cite{ARCSW,ACS}. 

Our in-medium results for the $\pi\pi$ amplitude $Im M_{\pi\pi}^{00}$,  
the spectral function $\eta_{00}$ of `$\sigma$' exchange and the 
corresponding central part of the coordinate $NN$ potential,  
eq.~(\ref{vnnsr}) (with the lower integration limit set to zero) 
are displayed in fig.~3 for both the BbS J\"ulich $\pi\pi$ model  
(upper panels) and its chirally improved version (lower panels).  
In order to provide a more realistic context for the attractive
contibution of the 2$\pi$ exchange, we supplemented this contribution
to the $NN$ central potential 
with a zero-width $\omega$ exchange characterized by 
\begin{equation} 
\eta_\omega(t')=\pi g_{NN\omega}^2 \delta(t'-m_\omega^2) 
\label{etaomega}
\end{equation} 
with $g_{NN\omega}^2/4\pi$=20, as in the Bonn potential~\cite{MHE}. 
(Due to the absence of a form factor, the potential which results
from eq.~(\ref{etaomega}) is substantially stronger, especially at
short range, than the $\omega$ exchange of the Bonn potential.)
The features found in $Im M_{\pi\pi}^{00}$ (left panels) are 
essentially reproduced by the `$\sigma$' spectral function (middle 
panels): in the BbS J\"ulich model these are an appreciable  
accumulation of strength around the 2$\pi$ threshold 
region evident already at half normal 
nuclear matter density ($\rho=0.5\rho_0$), together with a suppression 
in the higher energy range. At $\rho=\rho_0$ a strong $\pi\pi$ bound 
state emerges that absorbs the major part of the strength in the entire 
scalar-isoscalar channel. In terms of a one-boson exchange picture this 
corresponds to a $\sigma$-like particle with a mass of about 100~MeV! 
The consequences for the central part of the $NN$ potential are 
obvious: a 
dramatic increase in the attraction by more than an order of magnitude, 
even invading very short distances. This is at variance with the 
results obtained earlier~\cite{DKW} with the same model for the free 
$N\bar N \to \pi\pi$ interaction. The differences stem from the 
full off-shell treatment of the in-medium pion propagation in 
connection with the inclusion of the subthreshold region in the 
present work.  

With the chirally improved J\"ulich model the medium effects are much  
less pronounced (lower panels in fig.~3): the chiral constraints 
inhibit the 
development of $\pi\pi$ bound states at moderate densities---up to 
$\rho\simeq 1.6\rho_0$. However, a significant shift of strength to 
lower energies in both $Im M_{\pi\pi}^{00}$ and $\eta_{00}$ seems 
inevitable. This results in a smaller, but still appreciable increase 
in attraction 
in $V^C_{NN}(r)$ which, when incorporated into nuclear many-body 
calculations, is most likely incompatible with saturation. 
Thus we have to conclude that even though the chiral constraints 
on the $\pi\pi$ interaction substantially improve the situation, 
there remain important deficiencies in our microscopic description 
of the in-medium $NN$ potential. \\ 

One might first ask to what extent one can rely on the treatment of 
the pion propagation in nuclear matter.  The model of p-wave 
$\Delta N^{-1}$ and  $NN^{-1}$ excitations has been proven successful 
in different branches of nuclear pion physics~\cite{AEM,Oste}.  
However further refinements, which potentially reduce the rather 
pronounced softening of the in-medium pion dispersion relation,  
seem possible.  Among them are:
\begin{itemize} 
\item[$\bullet$] the inclusion of s-wave $\pi N$ interactions leading 
to a suppression of the pion self-energy~\cite{MSTV}. For 
low-momentum/energy pions especially such contributions  
might be significant, as the free $\pi N$ amplitude  
is also known to obey soft pion theorems; however, 
those will not be included  in our present study;  
\item[$\bullet$] Pauli blocking effects in the 
2$\pi$ propagator when both 
pions are simultaneously excited into a $NN^{-1}$ (or $\Delta N^{-1}$) 
mode, which amounts to taking into account certain exchange diagrams. 
A preliminary estimate of this effect indicated it to be rather 
small~\cite{APhD}; 
\item[$\bullet$] a density dependent increase of the short-range
correlation parameters $g'$~\cite{DiMu} (so far we used constant values
of $g'_{NN}$=0.8 and $g'_{N\Delta}$=$g'_{\Delta\Delta}$=0.5);
this point will be addressed at the end of the next section. 
\end{itemize} 

One might also ask if the $\pi\pi$ interaction kernel 
(pseudopotential) itself 
undergoes substantial medium modifications, in particular 
concerning the exchanged mesons (most importantly the $\rho$ meson). 
On the $\pi\pi$ level, this is just the analog of the modification 
of the `$\sigma$' exchange in the $NN$ potential discussed above. 
In the next section we will pursue such a possibility in terms of the 
Brown-Rho scaling hypothesis~\cite{BR91}, which asserts that, 
{\it e.g.}, vector meson masses decrease with increasing nucleon 
density. 
 
\subsection{Impact of Chiral Symmetry Restoration on the Central
$NN$ Potential}  
As has been proposed by Brown and Rho~\cite{BR91} the (partial) 
restoration of spontaneously broken chiral symmetry in hot/dense 
matter leads to an approximately universal decrease of most physical 
quantities like masses, coupling constants, form factor cutoffs, 
{\it etc}. 
The corresponding ``BR scaling'' conjecture as a function of 
temperature T and nuclear matter density $\rho$ reads:
\begin{equation}
\Phi(\rho,T)=\frac{m_V^*}{m_V}= \frac{m_\sigma^*}{m_\sigma}=
\frac{M_N^*}{M_N}=\frac{f_\pi^*}{f_\pi}
=\frac{\Lambda^*}{\Lambda} \ = \ldots etc. \ ,  
\label{BRscaling}
\end{equation}
where the asterisks indicate in-medium values. Due to the  Goldstone 
boson nature of the pion, its mass is excluded from this relation. The 
scale factor $\Phi(\rho,T)$ is governed by the decrease of the chiral 
quark condensate, conservatively estimated to be~\cite{AdBr}
\begin{equation}
\Phi(\rho,T)\simeq \left( \frac{ \langle 0|\bar qq|0\rangle (\rho,T)}
{\langle 0|\bar qq|0\rangle^{0}} \right)^{\frac{1}{3}} \ .  
\label{PHIrho}
\end{equation}
At zero temperature, and to leading order in density, the reduction of  
the quark condensate can be related to the pion-nucleon sigma term by 
\begin{equation}
\frac{\langle 0|\bar qq|0\rangle (\rho )}
{\langle 0|\bar qq|0\rangle^0} = 1-\frac{\Sigma_{\pi N}}
{m_\pi^2 f_\pi^2}  \ \rho \ .
\end{equation}
With $\Sigma_{\pi N}$=45~MeV one obtains $\Phi(\rho =\rho_0)$=0.87, 
which is compatible with QCD sum rule analyses of vector meson 
masses~\cite{HaLe} giving
\begin{equation}
\frac{m_V(\rho)}{m_V(0)}=1-C\frac{\rho}{\rho_0}
\label{linear}
\end{equation}
with $C$=0.18$\pm$0.06. In the following we will assume $\Phi(\rho)$ 
to drop linearly according to eq.~(\ref{linear}) with $C$=0.15. \\

In our present context, BR scaling enters on the level of the 
$\pi\pi$ interaction kernel $V_{\pi\pi}^{00}$ (pseudopotential) 
for both t-channel $\rho$ exchange (decreasing $m_\rho^*$, 
$\Lambda_\rho^*$) as well as $\pi\pi$ contact interactions 
(decreasing $f_\pi^*$, $m_\rho^*$), the latter appearing only in 
the chirally improved version. Since the KSFR relation~\cite{KSFR} 
is supposed to hold also in the medium, 
\begin{equation}
2g_{\pi\pi\rho}^2 (f_\pi^*)^2=(m_\rho^*)^2 \ ,  
\label{ksfr}
\end{equation}
the $\pi\pi\rho$ coupling constant is not affected.     
The many-body excitations ($NN^{-1}$ and $\Delta N^{-1}$) of the 
in-medium single-pion propagator are always evaluated in terms of 
effective baryon masses   
\begin{eqnarray}
M_N^* & = & M_N(1-0.2\rho/\rho_0)
\nonumber\\
M_\Delta^* & = & M_\Delta-(M_N-M_N^*) \ ,  
\end{eqnarray}
where the second relation accounts for the fact that the $\Delta$-$N$ 
mass difference in nuclei is not seen to alter very much. Within the 
uncertainties, this agrees with eq.~(\ref{linear}).   

Repeating the in-medium calculations as described in the 
previous section with BR scaling assumed, the following picture 
emerges (fig.~4): in the BbS J\"ulich model (upper panels) the 
dropping $\rho$ mass
leads to an increased attraction in the low-energy $\pi\pi$ interaction
which reinforces the shift of strength to (very) low energies in the
spectral function $\eta_{00}$. On the other hand, with the chirally 
improved version (lower panels), the simultaneous increase of 
repulsion in the 
$\pi\pi$ contact interactions (being proportional to $(f_\pi^*)^{-2}$)
counterbalances the (attractive) t-channel $\rho$ exchange. The 
corresponding central $NN$ potentials, supplemented with $\omega$ 
exchange according to eq.~(\ref{etaomega}) with $m_\omega^*$ from
eq.~(\ref{linear}), are displayed in the right panel of fig.~4: 
the BbS J\"ulich model still leads to an undesirable strong increase 
in attraction, whereas the chirally improved version 
results in a nice compensation between correlated 2$\pi$ (`$\sigma$')
and $\omega$ exchange. It is worthwhile to mention here that a very 
similar behavior is obtained if one replaces the correlated 2$\pi$ 
exchange by a sharp (zero-width) effective $\sigma$ exchange, as 
employed in earlier versions of the Bonn potential, and scales its 
mass $m_\sigma^*$ according to eq.~(\ref{linear}). In other words, 
the implementation of BR scaling in our dynamical model for 
correlated 2$\pi$ exchange reproduces an effective scaling of a 
fictitious $\sigma$ meson in a nontrivial way, thereby generating  
a considerable stabilization of the central $NN$ potential 
in nuclear matter.  

Note that once a universal scaling is established, its effects 
are not unexpected.  A simple, non-relativistic model with static 
central potentials of Yukawa form derived from $\sigma$ and $\omega$ 
exchange will saturate at {\it some} density if {\it all} effective  
masses---meson {\it and} nucleon---are scaled uniformly
downwards.  All that is required is that the $\omega$ potential
be stronger than the $\sigma$ potential at short range, and that
the $\omega$ mass be larger than the $\sigma$ mass.  We make no
claim that the $2\pi$ exchange interaction that we have calculated
(supplemented with $\omega$ exchange) will saturate at the correct
energy and density---only that it {\it will} saturate if BR scaling
is implemented in our framework. Quantitative predictions of the
saturation energy and density will clearly depend on the
constants of the model, such as the $\omega$ coupling constant
and form factor, the rate of change with density of the vector
meson masses (and related quantities, {\it cp.} eq.~(\ref{BRscaling})), 
as well as a possible density dependence of the Migdal 
parameters $g'$ entering the pion selfenergy, {\it e.g.}: 
\begin{itemize} 
\item[$\bullet$] when increasing the BR scaling factor from $C=0.15$ to 
$C=0.22$~\cite{FrRo}, and repeating our calculations as described 
above, the repulsive $\omega$ exchange dominates the attraction 
generated by correlated 2$\pi$ exchange: with increasing density
the minimum in $V_{NN}^C(r)$ gradually moves upwards in both energy and 
distance, eventually resulting in an entirely repulsive potential 
at densities around 2$\rho_0$;
\item[$\bullet$] on the other hand, a density dependence 
of the short-range correlation parameters is found to have only minor
impact on our results; in an exploratory calculation we chose 
\begin{eqnarray} 
g'_{NN}(\rho) & = & 0.6+0.2 \ \rho / \rho_0  
\nonumber\\
g'_{N\Delta}(\rho) & = & g'_{\Delta\Delta}(\rho) 
 =  0.33+0.17 \ \rho / \rho_0 \ , 
\end{eqnarray} 
which is a somewhat weaker increase than suggested in ref.~\cite{BBLW}, 
but coincides with our values used above at $\rho$=$\rho_0$. Using 
again $C=0.15$ in our meson exchange potentials, the resulting 
in-medium $NN$ potential from $\omega$ and correlated 
2$\pi$ exchange exhibits only marginal changes: compared to the 
results shown in the lower right panel 
of fig.~4, the long range attraction decreased by about 5\% 
at $\rho$=2$\rho_0$. 
\end{itemize} 

\section{Summary and Conclusions}
We have seen that a realistic model of the $\pi\pi$ interaction,
extrapolated by standard techniques to account for interactions with
a nuclear medium, immediately leads to serious difficulties in the 
form of a marked shift in (attractive) interaction strength in the
$\sigma$ channel to low energies.  This effect is directly traceable
to the polarization of the medium by the pion---through 
$\Delta$-nucleonhole and nucleon-nucleonhole excitations, 
which makes the sub-threshold energy region $E<2m_{\pi}$ 
accessible. This leads to an amplification of any attractive 
interaction strength in that energy range and results,
ultimately, in a $\pi\pi$ pairing instability.

The situation improves when certain constraints dictated by chiral 
symmetry are imposed on the effective $\pi\pi$ interaction: {\it i.e.}, 
employing a (broken) chirally symmetric interaction kernel as well 
as enforcing the correct chiral limit of the scattering length 
on the full amplitude.  
In that case the sub-threshold strength of the interaction is 
suppressed and the pairing instability does not set in until well above
normal nuclear matter density.  As we remarked in the introduction,
this shifts the problem; it does not cure it.  The in-medium 
nucleon-nucleon central interaction due to correlated 
2$\pi$ exchange still grows
increasingly attractive with density.  In the absence of mechanisms
which increase the short-range repulsion as the density of the medium 
increases, nuclear matter will fail to saturate.

We have already enumerated some of the possible sources for this
additional repulsion.  It is also possible---indeed, almost 
certain---that a fuller treatment of the 
chiral dynamics in the medium would 
result in modifications of the effective $\pi\pi$ interaction beyond
our rather simple enforcement of the chiral constraints.
For example,  our way of imposing the scattering length constraint 
guarantees only that the first
two terms in the momentum expansion of the $\pi\pi$ amplitude are 
correct; higher-order terms are generated solely by the fit of the
model to the free $\pi\pi$ data and are probably not consistent with
chiral symmetry~\cite{Wirz}. However, such a calculation over the broad
energy and density ranges required for our purposes is far beyond our 
means, so that, for the present, we have to limit our investigations
to what can be done with the best available effective model.

Our search for an effective mechanism for providing the additional
repulsion needed to stabilize nuclear matter led us to consider the
Brown-Rho scaling scenario~\cite{BR91} in which, {\it e.g.}, vector 
meson masses decrease with increasing
density.  We found that a consistent scaling of masses, coupling
constants, {\it etc.} 
does, indeed, lead to sufficient repulsion in the nucleon-nucleon
interaction at higher densities which
counteracts the growing attraction due to $\sigma$ channel 2$\pi$
exchange.  While it is clear that BR scaling {\em is} an efficient 
mechanism---that is, in our approach it works to stabilize nuclear 
matter, its dynamical origins are not clear~\cite{Chan}. Thus, 
we are left with at least two important questions:
\begin{itemize}
\item[$\bullet$]Do effects which we have so far neglected in our 
treatment of pion propagation sufficiently ``stiffen'' the in-medium 
pion dispersion relation and thereby provide the repulsion necessary 
for the saturation of nuclear matter?
\item[$\bullet$]Is BR scaling contained in a more complete conventional 
approach to the dynamics of pions in nuclear matter, or is it 
a consequence of QCD in the 
intermediate energy range which cannot be derived from effective 
interactions involving a limited set of mesons and nucleons?
\end{itemize}

These two questions are, in fact, facets of the same question.  
If the answer to the first is `yes', then we would expect the 
inclusion of the neglected processes
to produce effects which resemble BR scaling.  In that case, 
the scaling would be simply an economical way to express these effects.  
If the answer is `no', then it may be necessary to promote the 
BR scenario from an hypothesis to a rule
for constructing meson exchange models of hadronic interactions at 
non-zero density, and to seek its origin in QCD at a different 
level from effective meson exchange interactions.  The answer to 
these questions appears to be worth some effort.
\vskip1cm
 
\centerline {\bf ACKNOWLEDGMENTS}
We are grateful for productive conversations with G.E. Brown, J. Speth 
and  H.C. Kim. 
One of us (JWD) wishes especially to thank Prof. Speth 
for his hospitality and support during JWD's visits to the 
Forschungszentrum J\"ulich.  One of us (RR) acknowledges financial 
support from the Alexander-von-Humboldt foundation as a Feodor-Lynen 
fellow. This work was supported in part by the National Science 
Foundation under Grant No. NSF PHY94-21309 and by the U.S. Department 
of Energy under Grant No. DE-FG02-88ER40388.

\pagebreak
 
\begin{center} 
{\large \sl \bf Figure Captions}
\end{center}
\vspace{0.5cm}

\begin{itemize}
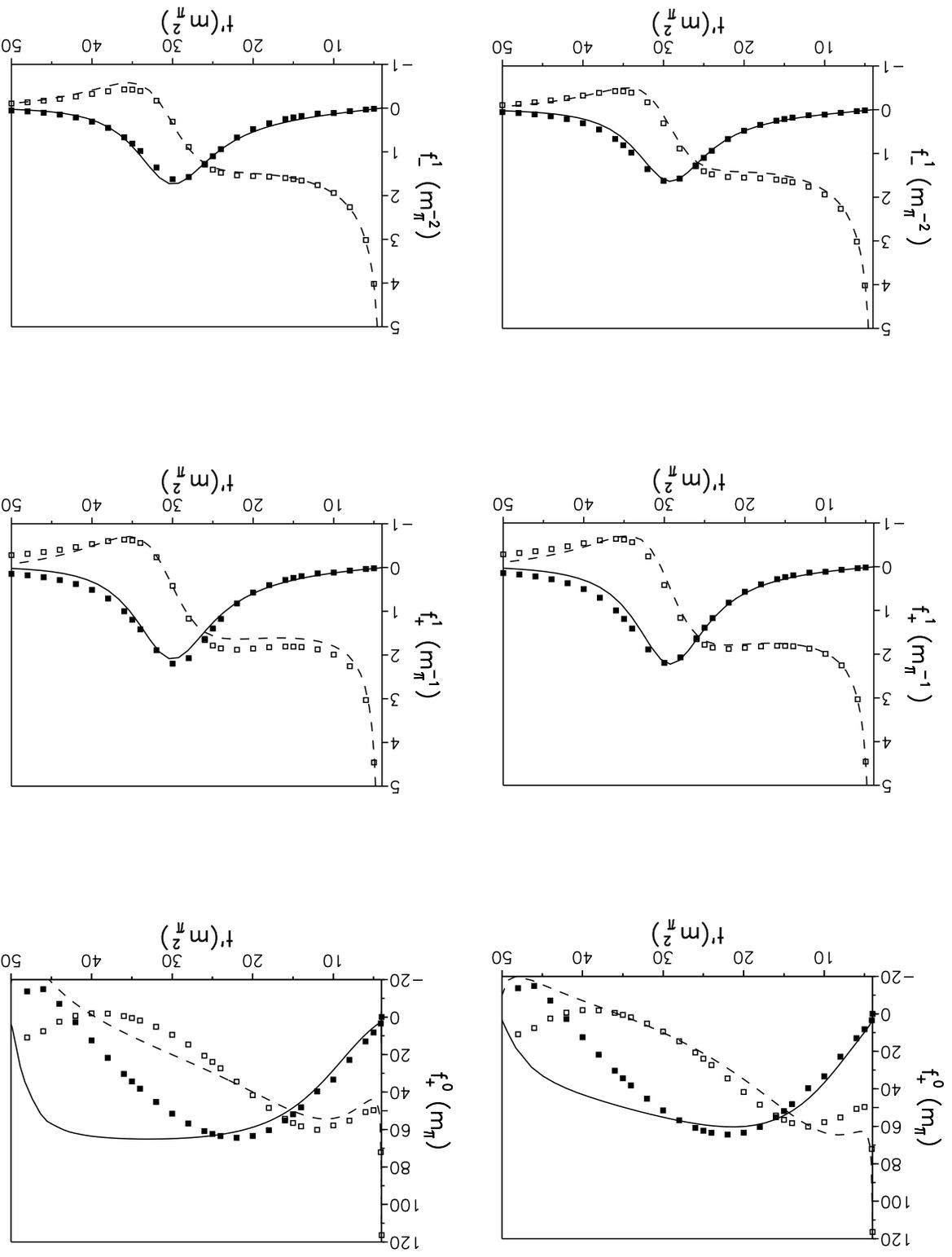
\item[{\bf Figure 1}:] 
Our fit to the quasi-empirical Frazer Fulco amplitudes~\cite{Hoeh} in 
free space (open squares: real part; solid squares: imaginary part) 
employing two different models for the underlying $\pi\pi$ interaction, 
namely the BbS J\"ulich model~\cite{PHS,SHSPD} (left column) and a 
chirally improved version~\cite{RDW} (right column).  
 
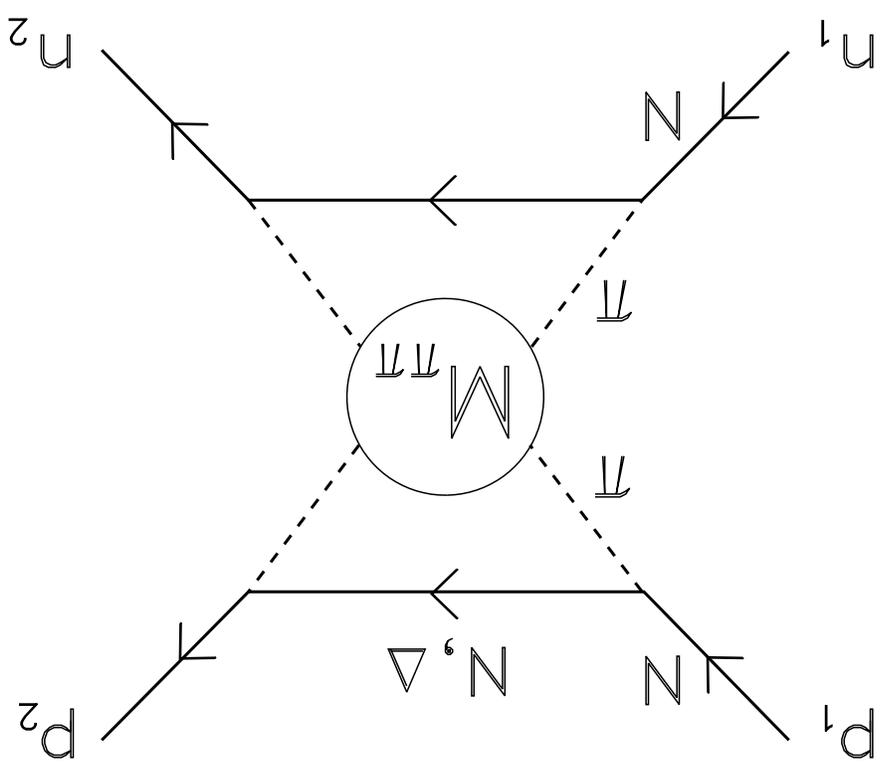
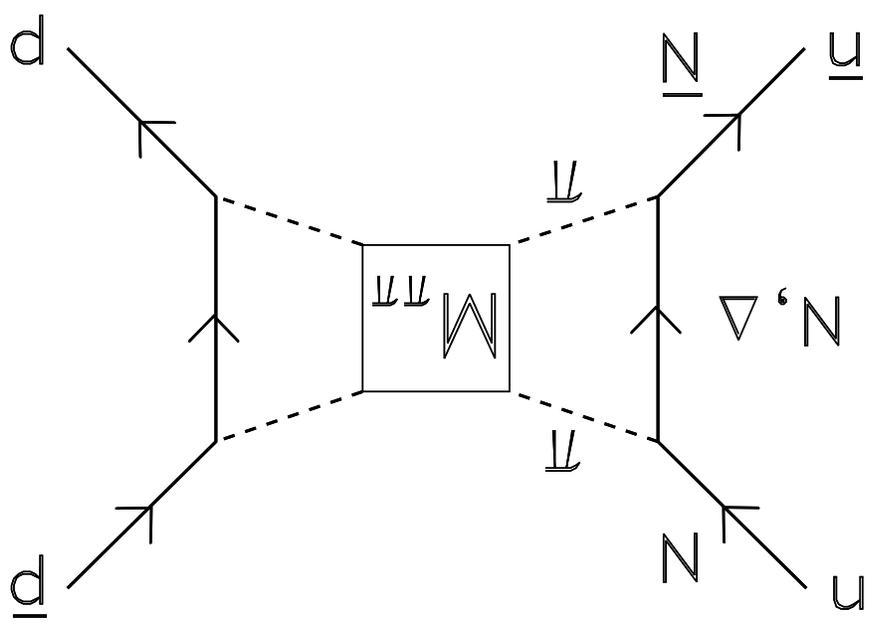
\item[{\bf Figure 2:}]
Our meson exchange model for the correlated two-pion exchange 
contribution to the nucleon-nucleon potential; \\ 
upper panel: microscopic model for $N\bar N \to \pi\pi \to N\bar N$ 
scattering with in- and outgoing $N$,$\bar N$ 4-momenta 
$n=(E_n,\vec n)$, $\bar n =(E_{\bar n},-\vec n)$ and $p=(E_p,\vec p)$, 
$\bar p =(E_{\bar p},-\vec p)$, respectively; \\ 
lower panel: corresponding $N$-$N$ potential for in- and 
outgoing nucleon
4-momenta $p_1=(E_{p_1},\vec k)$, $n_1=(E_{n_1},-\vec k)$ and 
$p_2=(E_{p_2},\vec k')$, $n_2=(E_{n_2},-\vec k')$, respectively. \\
The time direction in both diagrams is pointing from left to right.  
 
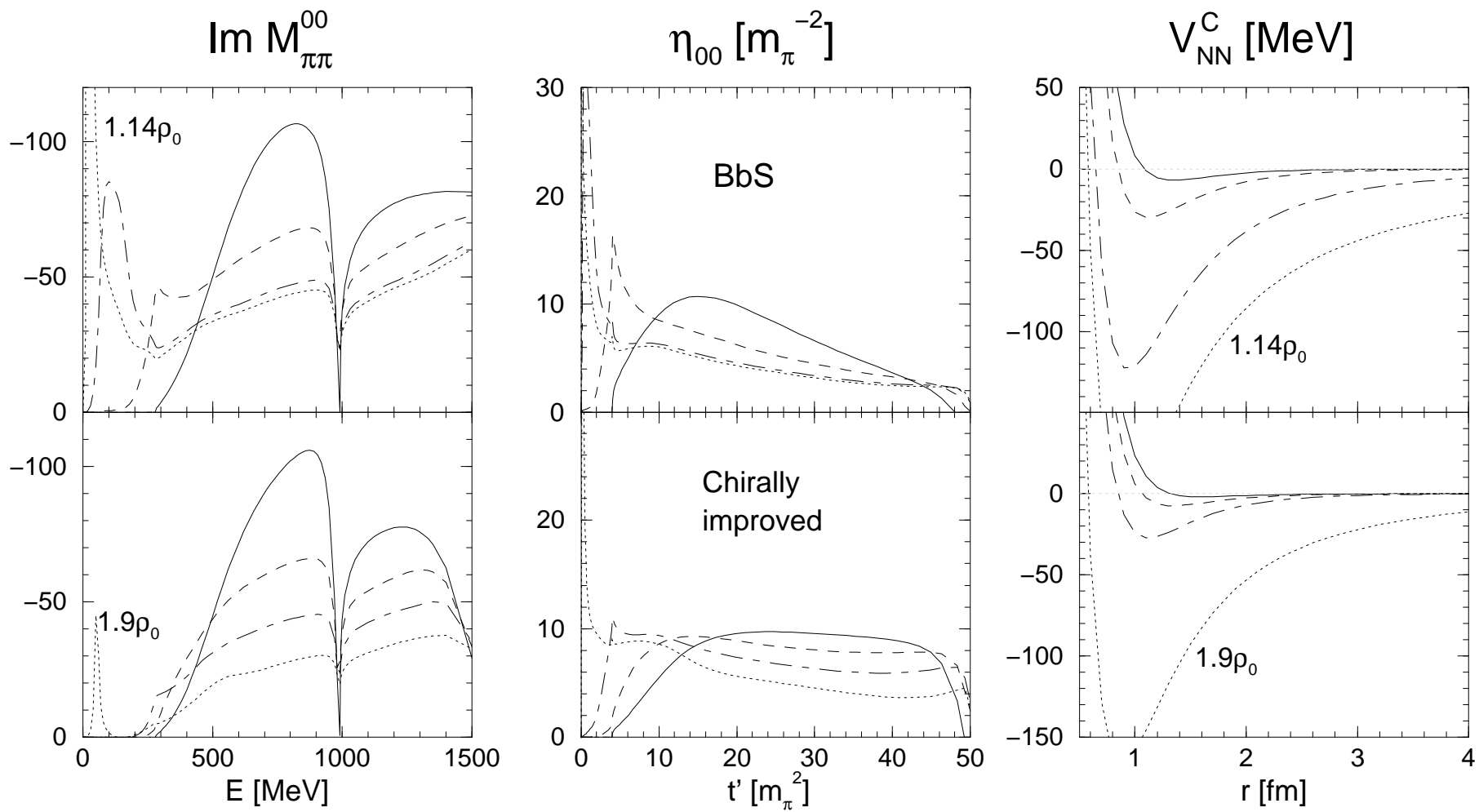
\item[{\bf Figure 3:}]  
Imaginary part of the $\pi\pi$ scattering amplitude (left panels) and 
$N\bar N$ spectral function (middle panels) in the $\sigma$ channel 
($JI$=00) as well as the corresponding central $N$-$N$ potential 
(right panels), supplemented with (zero-width) $\omega$ exchange, 
at various  nuclear matter densities (full lines: free space; dashed 
lines: $\rho$/$\rho_0$=0.5; dashed-dotted lines: $\rho$/$\rho_0$=1; 
dotted lines: $\rho$/$\rho_0$=1.14 in the upper panels and 
$\rho$/$\rho_0$=1.9 in the lower panels); the upper panels show the 
results for the BbS J\"ulich model, the lower ones for the 
chirally improved version. 
 
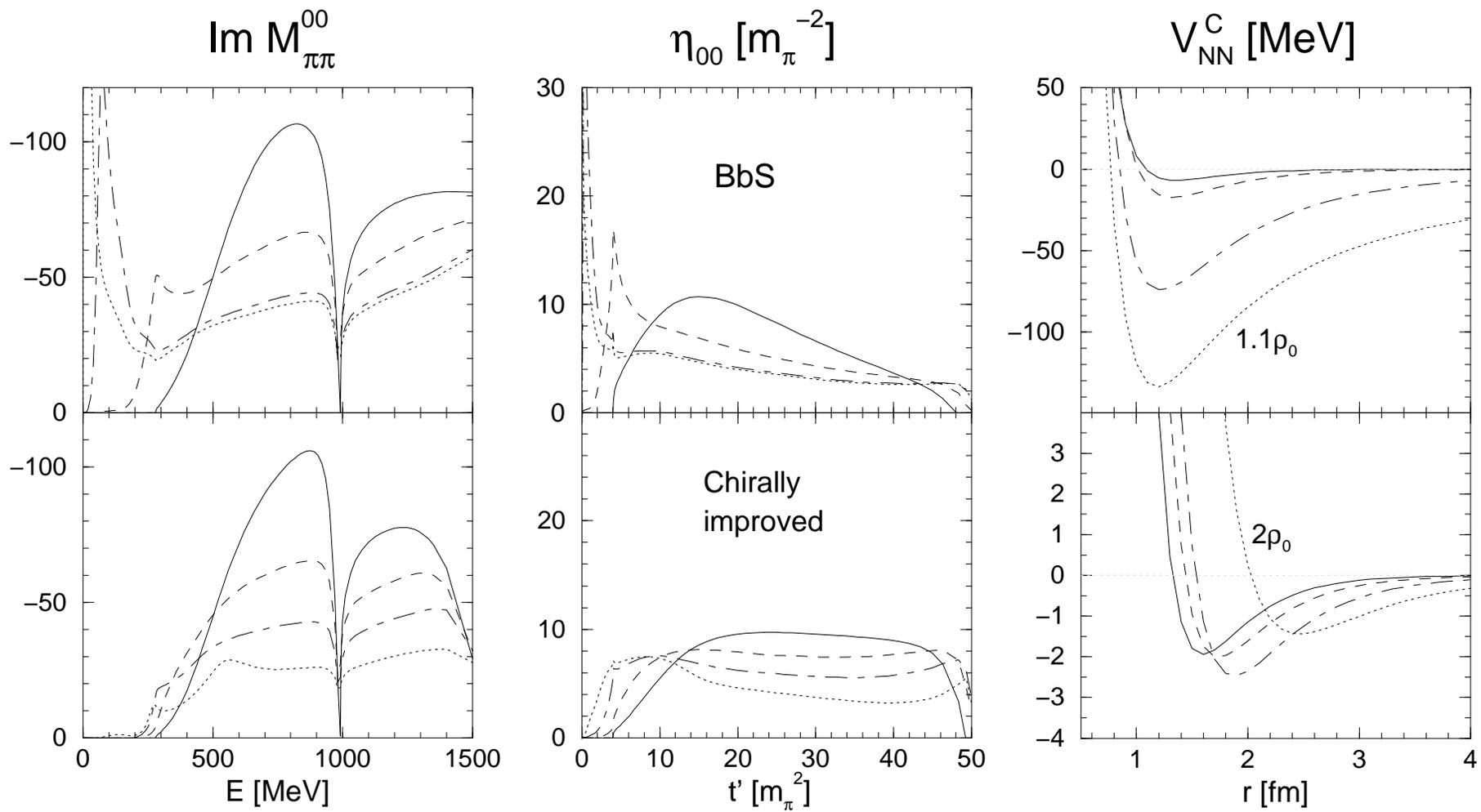
\item[{\bf Figure 4:}] 
Same as fig.~3, but including the BR scaling as described in the text 
with  a scaling factor $C$=0.15 ({\it cp.} 
eqs.~(\ref{BRscaling}),~(\ref{linear})).  
 
\end{itemize}


\begin{thebibliography}{99}
\bibitem{Wirz} 
We wish to thank A. Wirzba for pointing this out to us.
\end{thebibliography}
\end{document}